# Low-loss silicon nitride photonic ICs for single-photon applications

**Kirill A. Buzaverov,**[1,2] **Aleksandr S. Baburin,**[1,2] **Evgeny V. Sergeev,**[1] **Sergey S. Avdeev,**[1] **Evgeniy S. Lotkov,**[1] **Mihail Andronik,**[1] **Victoria E. Stukalova,**[1] **Dmitry A. Baklykov,**[1] **Ivan V. Dyakonov,**[3] **Nikolay N. Skryabin,**[3] **Mikhail Yu. Saygin,**[3] **Sergey P. Kulik,**[3] **Ilya A. Ryzhikov,**[1] **and Ilya A. Rodionov**[1,2,*]

[1]*FMN Laboratory, Bauman Moscow State Technical University, Moscow 105005, Russia*
[2]*Dukhov Research Institute of Automatics (VNIIA), Moscow 127055, Russia*
[3]*Quantum Technology Center, Faculty of Physics, Lomonosov Moscow State University, Moscow 119991, Russia*
*\*irodionov@bmstu.ru*

**Abstract:** Low-loss photonic integrated circuits (PICs) are the key elements in future quantum technologies, nonlinear photonics and neural networks. The low-loss photonic circuits technology targeting C-band application is well established across multi-project wafer (MPW) fabs, whereas near-infrared (NIR) PICs suitable for the state-of-the-art single-photon sources are still underdeveloped. Here, we report the labs-scale process optimization and optical characterization of low-loss tunable photonic integrated circuits for single-photon applications. We demonstrate the lowest propagation losses to the date (as low as 0.55 dB/cm at 925 nm wavelength) in single-mode silicon nitride submicron waveguides (220x550 nm). This performance is achieved due to advanced e-beam lithography and inductively coupled plasma reactive ion etching steps which yields waveguides vertical sidewalls with down to 0.85 nm sidewall roughness. These results provide a chip-scale low-loss PIC platform that could be even further improved with high quality $SiO_2$ cladding, chemical-mechanical polishing and multistep annealing for extra-strict single-photon applications.

## 1. Introduction

Photonic integrated circuits (PICs) are under focused attention due to their high potential for future applications in telecom and datacom [1-3], LiDAR [3, 4], biophotonics [5, 6], nonlinear photonics [8, 9], neural networks [10-12] and quantum technologies [13-19]. In a wide variety of material platforms suitable for photonic integration, scalable and power-efficient solutions nowadays are driven by Si [20, 21], InP [22, 23] and $Si_3N_4$ [24, 25]. Despite current technology limitations on active elements and modulation speeds, silicon nitride platform exploits the advantage of passive components with lowest losses in a wide wavelength range from visible to mid-IR in integrated circuits [26]. Using hybrid assembly, flip-chip integration and wafer bonding with $A_3B_5$ or SOI platforms [27, 28], silicon nitride photonic platform could be equipped with all the necessary tools for fully integrated optical signal processing.

Ultra-low (<0.01 dB/cm) propagation losses are the key to high-efficiency devices on PICs. To achieve such low values, three major origins of losses have to be considered. The origins are absorption, radiation and scattering losses. Using high quality wet and dry oxidation with LPCVD $SiO_2$ and stoichiometric $Si_3N_4$ films together with high temperature annealing, absorption losses could be lowered down to less than 0.1 dB/m [29]. With advanced bends optimization techniques, radiation losses due to mode mismatch can be reduced to 0.012 dB/90º bend values [30, 31]. Thus, scattering losses from bottom, sidewalls and upper surfaces roughness are the main source of light attenuation in waveguides. Compared to upper and bottom surfaces roughness reduction methods (usually chemical-mechanical polishing [32]), reduction of waveguide sidewall roughness still remains the primary technological challenge [29, 33].

There are three main fabrication processes proposed for effective minimization of propagation losses in silicon nitride photonic integrated circuits. The first one is the photonic Damascene reflow process, which allows to fabricate circuits with crack-free silicon nitride films up to 1.5 μm thickness [33]. Using stress management, $SiO_2$ preform reflow and high precision chemical-mechanical planarization, silicon nitride microresonators with quality factor up to $32·10^6$ (equal to 1 dB/m at 1550 nm wavelength) were fabricated [34]. However, extremely high annealing temperatures lead to silicon diffusion in $SiO_2$, degrading optical properties of both materials. Thicker layers of silicon dioxide are required to prevent absorption losses [34]. It was also found out that $SiO_2$ preform reflow may introduce transition metals (Cr, Fe, Cu) impurities and enhance their diffusive redistribution [35], which causes wavelength-independent absorption losses. The above limitations together with high requirements for chemical-mechanical planarization constrain the application of the Damascene reflow process in R&D laboratories for experimental and small-scale batches of devices.

The second one is the classic subtractive process. It is an alternative to the photonic Damascene fabrication process [36]. Proper optimization of individual fabrication steps (mainly lithography and etching steps) results in circuits with low losses (microresonators with quality factor up to $70·10^6$, equivalent to 0.4 dB/m at 1550 nm wavelength). The technology utilizes $Si_3N_4$ thicknesses up to 1 μm with multistep LPCVD deposition process without long extremely high-temperature annealing and high precision chemical-mechanical planarization steps.

The third one is the silicon nitride TriPleX® waveguide technology [37]. It is a variation of subtractive process for fabrication of ultralow-loss silicon nitride waveguides. Using waveguides with high-aspect ratio core (single or double stripe, up to 100 nm thickness, up to 10 um width) record propagation losses in silicon nitride based photonic circuits are achieved (microresonators with quality factor up to $422·10^6$, equivalent to 0.06 dB/m at 1550 nm wavelength [37]). The approach is characterized be very low mode confinement and weak interaction of mode with waveguide surfaces roughness [26]. However, low confinement requires very thick lower $SiO_2$ layers (up to 15 μm) and also imposes restriction on critical bending radius of waveguides and microresonators (up to 1 cm), which limits scalability of the technology [38].

The above results are obtained at infrared C-band which is actively used in telecommunications, LiDAR, and quantum technologies. However, there are many applications where ultralow losses at wavelengths from 900 to 940 nm are crucial [17, 18, 39, 40]. In this wavelength range single-mode waveguides have to be significantly narrower down to 600 nm width with upper $SiO_2$ cladding (versus several microns width at 1550 nm wavelength), leading to higher interaction of waveguide mode with inhomogeneities and surfaces roughness, and thus to higher propagation losses. The lowest propagation losses of 0.6-1.5 dB/cm have been reached in silicon nitride waveguides with the classic subtractive fabrication process [41-43]. With silicon nitride TriPleX® technology the first 12-mode quantum photonic processor was fabricated demonstrating propagation losses less than 0.3 dB/cm at 940 nm operation wavelength [17]. In [18] silicon nitride PICs with on-chip quantum single-photon source are fabricated, reaching the lowest propagation losses to date – 0.01 dB/cm at 920 nm wavelength. Despite ultralow losses, authors declared very low single-photon coupling efficiency due to low mode confinement. To achieve higher coupling efficiency with medium and high mode localization in silicon nitride waveguides, comparable to the C-band, further improvement of the fabrication processes is required with an emphasis on studying the effect of their parameters on the reduction of surface roughness.

In this work, we report on the fabrication and optical characterization of low-loss silicon nitride photonic integrated circuits, with propagation losses less than 0.6 dB/cm at 925 nm wavelength. We perform the first in-depth, to the best of our knowledge, study of the influence of e-beam lithography and reactive ion etching parameters on the roughness of waveguide sidewalls. The study formulates rules to fabricate low-loss waveguides with smooth sidewalls. Using classic subtractive fabrication process based on optimized e-beam lithography and inductively coupled plasma reactive ion etching (ICP-RIE), we fabricated 550 nm-width single-mode submicron waveguides with sidewall roughness less than 1 nm and near 90º sidewalls angle, directional couplers with a gap down to 100 nm, grating and 120 nm-width taper couplers with insertion losses less than 8 dB and 4 dB, respectively. In this paper we propose the numerical solution for Payne-Lacey model, which allows accurate modeling of propagation losses in waveguides taking into account all three standard sidewall roughness parameters (standard deviation, correlation length and roughness exponent).

The paper is organized as follows. Section 2 presents modern methods for sidewall roughness measurement and explains the measurement technique used in this work. Section 3 presents the fabrication process of low-loss silicon nitride photonic integrated circuits with in-depth analysis of e-beam lithography and inductively coupled plasma reactive ion etching processes. Section 4 presents up-to-date approaches for propagation losses modeling in waveguides with proposed numerical solution taking into account roughness exponent in Payne-Lacey model. Finally, section 5 presents optical characterization of photonic integrated circuits elements – propagation and coupling losses are measured with "cut-back" technique.

## 2. Sidewall roughness measurement

There are two commonly used methods for sidewall roughness measurements: specialized atomic force microscopy (AFM) and scanning electron microscopy (SEM). The first one with rotated scanning axis and customized AFM tips allows to carefully measure sidewall roughness with atomic-scale sensitivity and resolution approximately 0.1 nm [44]. However, this technique is very expensive, time consuming and nonstandard for most R&D laboratories. Recently, the method for direct sidewall roughness measurement of waveguides with standard AFM was presented. It is based on Bosch deep silicon etching to fabricate a tall and thin silicon pillar with waveguide on top [45]. The above techniques require specialized metrology tools or high precision fabrication process with optimization of several photolithography and deep silicon etching steps. We choose scanning electron microscopy for sidewall roughness measurement, as this is a straightforward and convenient method. It is less accurate compared to AFM due to noise, aberrations, and charge accumulation [46]. However, for SEM based sidewall roughness measurement, the only requirement is algorithm for edge detection and roughness parameters extraction (Fig. 1).

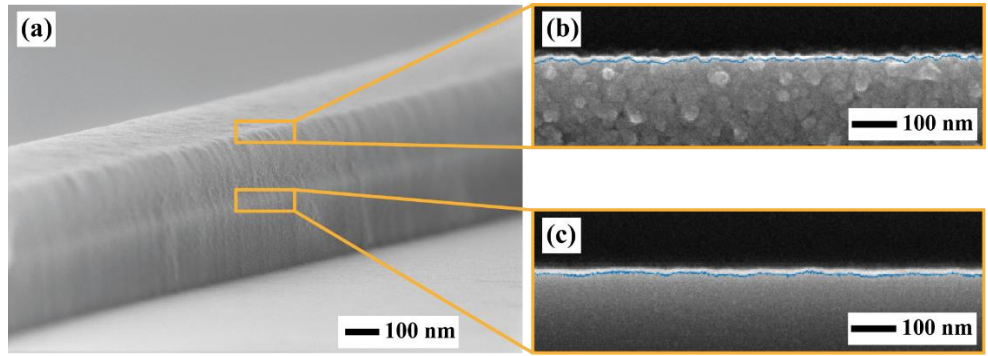

Fig. 1. (a) SEM image of waveguide sidewall (with resist on top). (b) Detected edge on SEM image of e-beam resist (blue line). (c) Detected edge on SEM image of etched silicon nitride waveguide (blue line).

In this work, we develop the sidewall roughness measurement algorithm using MATLAB software and methodology from [47-50]. It includes waveguide edge detection based on SEM image which requires no image filtration and frequency-domain analysis of obtained data for eliminating noise and extraction of three main roughness parameters – root-mean-square (RMS) roughness $\sigma$, correlation length $\xi$ and roughness exponent $H$ (more information about roughness parameters measurement could be found in Supplementary materials). Figure 1 demonstrates waveguide edges detection from SEM images with developed algorithm.

## 3. Fabrication of low-loss silicon nitride photonic integrated circuits

Fabrication process of low-loss silicon nitride photonic integrate circuits with light coupling through grating couplers is shown in Figure 2. Although grating couplers provide access to any location in the PIC and do not require complex post-fabrication processing, they provide coupling efficiency up to 60 % [51]. For higher coupling efficiency, which is critical for single-photon applications, much more complicated fabrication process and optimized coupling structure can be used. This purpose requires developing optical grade quality multistep deep reactive ion etching techniques to form precise edge of waveguides for coupling to photonic integrated circuit. [36, 52-54].

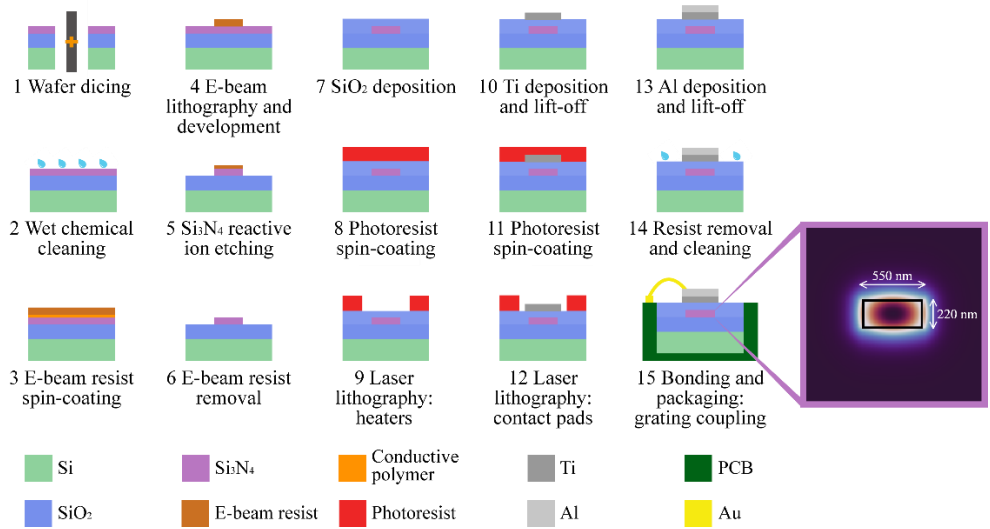

Fig. 2. The fabrication process of silicon nitride photonic integrated circuits. Inset: Mode simulation of a single-mode 220 nm tall and 550 nm wide waveguide at 925 nm wavelength.

We use commercially available 525 um thick silicon wafers with 2.5 um thickness of wet $SiO_2$ and 220 nm thickness of stoichiometric LPCVD $Si_3N_4$ from Silicon Materials Inc. (USA). In our circuits we use single-mode silicon nitride waveguides with 550 nm width and 220 nm height (Fig. 2). The wafers are diced into standard 25x25 mm$^2$ dies, which are then cleaned from organic, mechanical and metal residues. After that, e-beam lithography and inductively coupled plasma reactive ion etching are used for waveguides patterning with subsequent upper $SiO_2$ cladding deposition. For tunable beam splitters (Mach-Zehnder interferometers with thermo-optic phase shifters) we fabricate heaters and wiring with laser lithography and lift-off processes. After final cleaning, we bond the fabricated photonic integrated circuit to printed circuit board (PCB) and package for further testing. In next sections, we focus on in-depth study and optimization of e-beam lithography and ICP-RIE steps of the above fabrication process.

### 3.1 Electron-beam lithography and development

Electron-beam lithography was done using a beam with current of 400 pA to direct-write our circuits into a MaN-2403 negative resist. While direct writing by e-beam is a powerful and flexible method for fabricating features with the critical dimension less than 100 nm, its drawback is appearance of stitches at working fields boundaries due to lens aberrations and motorized stage instability. Also, due to statistical errors (fluctuations of beam current, vibrations, jitter), sidewall roughness of features could be increased [55, 56]. Using substrates with dielectric materials (SiO$_2$, Si$_3$N$_4$), accumulation of charge leads to increased sidewall roughness and decreased resolution [57]. These limitations lead to topology defects (Fig. 3) and dramatic increase in PICs loss, thus requiring the optimization of e-beam lithography parameters.

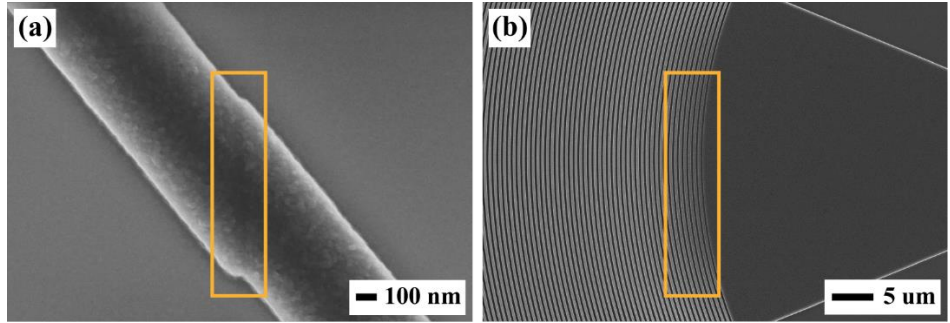

Fig. 3. SEM images of defects in e-beam lithography patterned structures: (a) Stitching error in waveguide, leading to increase in propagation losses. (b) Overexposure of resist lines, leading to change in dimensions and increase in roughness.

For stitching error reduction, a multipass e-beam writing and conductive polymer are the most effective techniques [58, 59]. In this work we optimize number of writing passes, the size of working field (WF) and conductive polymer development temperature for stitching error and resist sidewalls roughness reduction. To estimate the stitching error, more than 200 measurements for each specimen were carried out. Sidewall roughness was evaluated from 18 SEM images of waveguide edges with a length of 1 μm. Figure 4 shows stitching error and resist sidewall RMS roughness dependence versus different exposure parameters (number of writing passes and WF size).

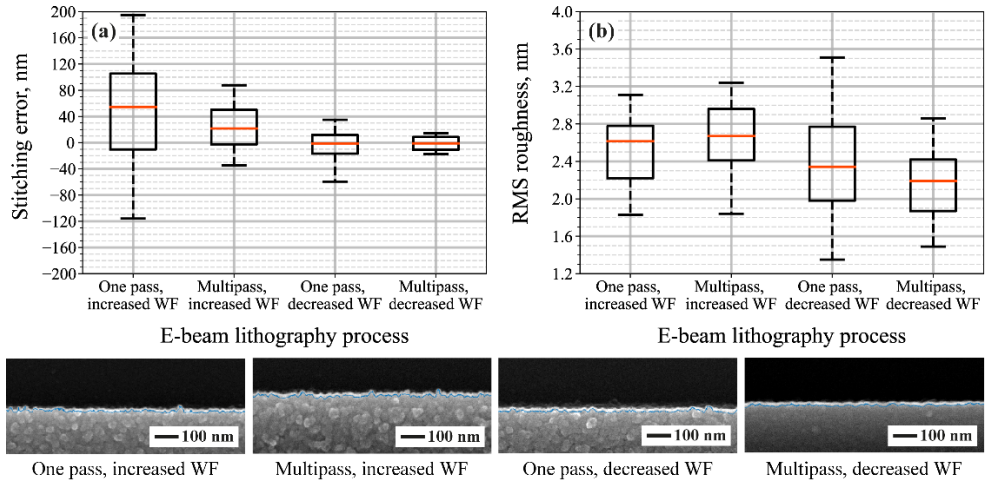

Fig. 4. Exposure parameters versus: (a) Stitching error. (b) RMS roughness. Average values are shown with solid orange line. Black solid boxes show interquartile range, within which 50% of measured values fall. Dashed lines show measured values outside the interquartile range with minimum and maximum values (horizontal lines). Inset: SEM images of resist sidewalls with detected edges (blue line), fabricated with different exposure parameters (number of writing passes and WF size).

Difference between one pass and multipass writing is observed in reducing the average stitching error modulo from 55.17 ± 11.41 nm to 21.90 ± 6.00 nm. Decreasing working field allows further reduction of the stitching error modulo down to 1.04 ± 1.35 nm. Statistical exposure errors, such as current fluctuations, jitter and beam drift, as well as mechanical vibrations and positioning errors are reduced with above techniques, leading to a significant reduction of stitching error. Also, multipass writing lithography and decreased working field allowed to reduce average RMS roughness from 2.63 ± 0.34 nm to 2.15 ± 0.24 nm. Multipass writing reduces the beam dwell time, thereby reducing charge accumulation and primary electron beam deflection, which has a positive effect on the roughness. The choice of a reduced working field reduces the distortion of the electron beam as it deviates from the optical axis, which also improving RMS sidewall roughness of e-beam resist.

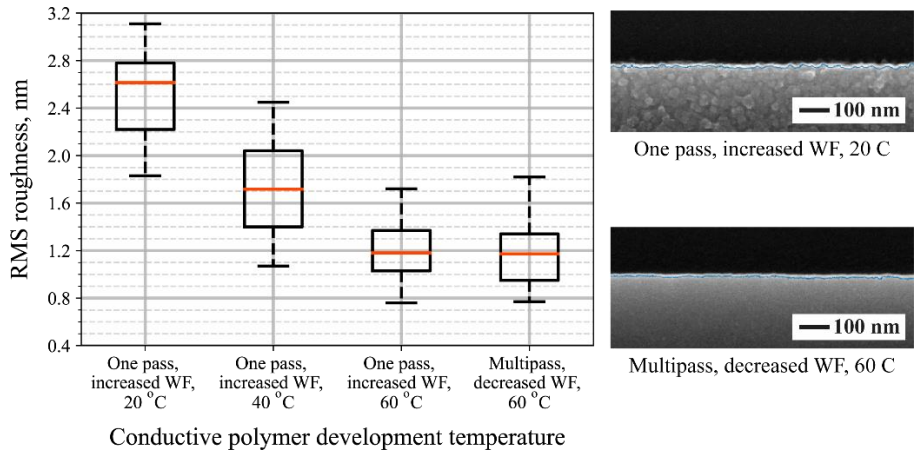

Fig. 5. RMS roughness dependence on conductive polymer development temperature. Average values are shown with solid orange line. Black solid boxes show interquartile range, within which 50% of measured values fall. Dashed lines show measured values outside the interquartile range with minimum and maximum values (horizontal lines). Inset: SEM images of resist sidewalls with detected edges (blue line), fabricated with different conductive polymer development temperature and exposure parameters (number of writing passes and WF size).

Additionally, we studied RMS roughness changing from conductive polymer development temperature. In the experiments, we varied the development temperature in deionized water from 20 ºC to 60 ºC. As shown in Figure 5, the temperature change reduces the average RMS roughness from 2.63 ± 0.34 nm to as low as 1.17 ± 0.13 nm with one pass exposure with increased working field. We assume that conductive polymer development in warm deionized water leads to it complete removal from resist surface, thereby smoothing resist sidewalls. Multipass lithography with reduced working field and warm conductive polymer development significantly reduces stitching error and RMS sidewall roughness of e-beam resist, thus can be used for smooth and precise mask layers fabrication.

*3.2 Inductively coupled plasma reactive ion etching*

Dry etching is one of the key technology issues for nanofabrication, capable to provide smooth surfaces and anisotropic etch profiles, with accurate control of process parameters [60]. In the reactive ion etching experiments RMS sidewall roughness, sidewall angle and $Si_3N_4$ to resist etching selectivity were optimized by varying RF bias power (HF), coil power (ICP), platen temperature (T), working pressure (p), and composition of gas mixture. In this case we first check the influence of parameters that directly affect the energy of particles in plasma. Figure 6 shows experimental results for RMS sidewall roughness, sidewall angle and etching selectivity (defined as the ratio of the etch rate of $Si_3N_4$ to the etch rate of resist) with $Si_3N_4$ and resist etching rates versus HF power for different etching regimes. HF power reduction from 250 W to 75 W shows a downwards trend for RMS roughness with sidewall angle improvement (blue and green line in Fig. 6a), due to lowering the energy of ions in discharge. Further reduction to 40 W leads to enhanced polymerization with formation of thin C(H,F)N passivation film [61] and increase in roughness (purple and orange line in Fig. 6a). We should note that more active isotropic etching occurs with higher platen temperature (purple line in Fig. 6a), leading to lateral etching of sidewalls and passivation film partial removal. This effect provides lower roughness and better angle of sidewalls compared to etching at lower temperatures. As shown in Figure 6b, change in platen power simultaneously changes etching rates of $Si_3N_4$ and resist with no change in etching selectivity for most regimes. One can see, that etching selectivity reaches higher values for lower platen temperature and HF power, possible due to higher plasma resistance of e-beam resist. Such an effect can stem from the formation of denser chemical bonds in the resist structure.

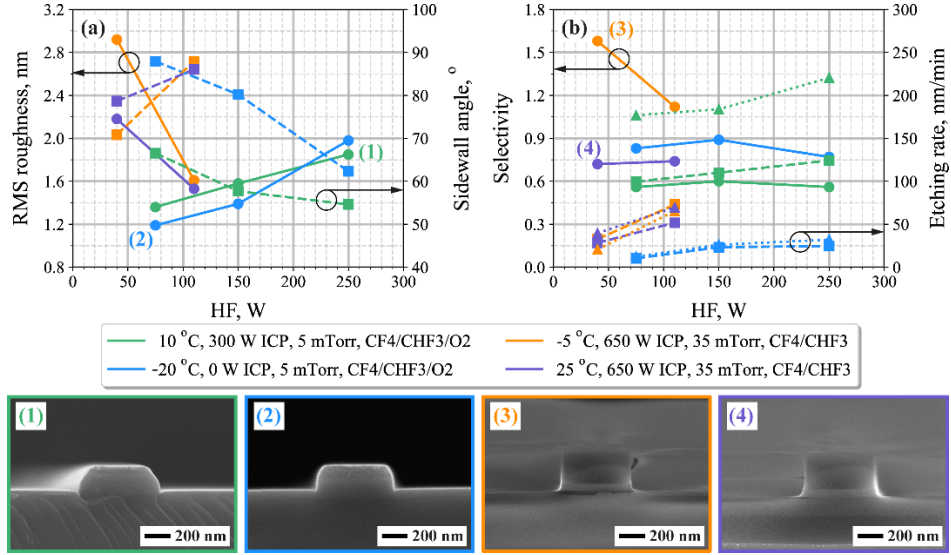

Fig. 6. Waveguides parameters versus platen power: (a) RMS sidewall roughness (solid lines) and sidewalls angle of inclination (dashed lines). (b) Etching selectivity (solid lines), $Si_3N_4$ and resist etching rates (dashed and dotted lines, respectively). Cross-section of waveguides, fabricated with various etching processes are shown on SEM images.

Results of optimization parameters on ICP power for various etch regimes are shown in Figure 7. Increasing ICP power from 300 W to 700 W at lower working pressures leads to higher RMS roughness with improved sidewalls angle due to higher energy ions and higher anisotropy of etching (green line in Fig. 7a). We observed strong damage of resist profile together with notching effect at ICP power above 500 W (SEM images inset in Fig. 7). At higher working pressures an opposite trend occurs with increase in roughness at lower ICP power (purple and orange lines in Fig. 7a) due to enhanced sidewalls passivation. Similar to selectivity dependence on HF power, there is no significant change in selectivity at lower pressure (Fig. 7b). At higher pressure selectivity decreases with higher ICP power as resist etch rate is higher compared to $Si_3N_4$ etch rate in discharge with high energy ions.

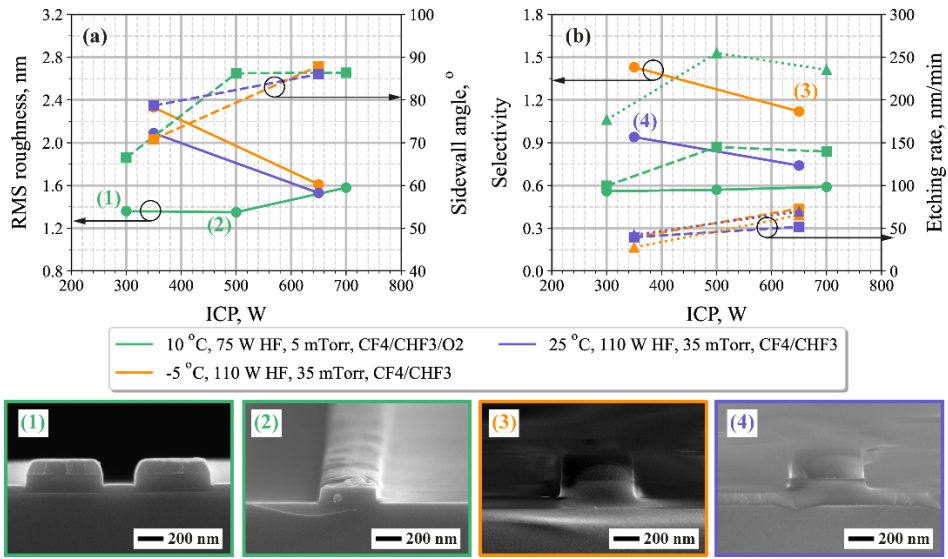

Fig. 7. Waveguide parameters versus on coil power: (a) RMS sidewall roughness (solid lines) and sidewalls angle of inclination (dashed lines). (b) Etching selectivity (solid lines), $Si_3N_4$ and resist etching rates (dashed and dotted lines, respectively). Cross-section of waveguides, fabricated with various etching processes are shown on SEM images.

At the next step platen temperature was varied (Fig. 8). Etching with lower temperature causes sidewalls with higher roughness and worse sidewall angle for all regimes. At lower pressure this effect is associated with higher temperature gradient and stronger ion bombardment. At higher pressure, deposited passivation film becomes more plasma resistant at low temperature and cause rough sidewalls formation. The selectivity also increases with low temperature etching for all regimes. As mentioned earlier, this effect may be associated with a change in the resist structure during cooling, which is confirmed by a decrease in the resist etching rate.

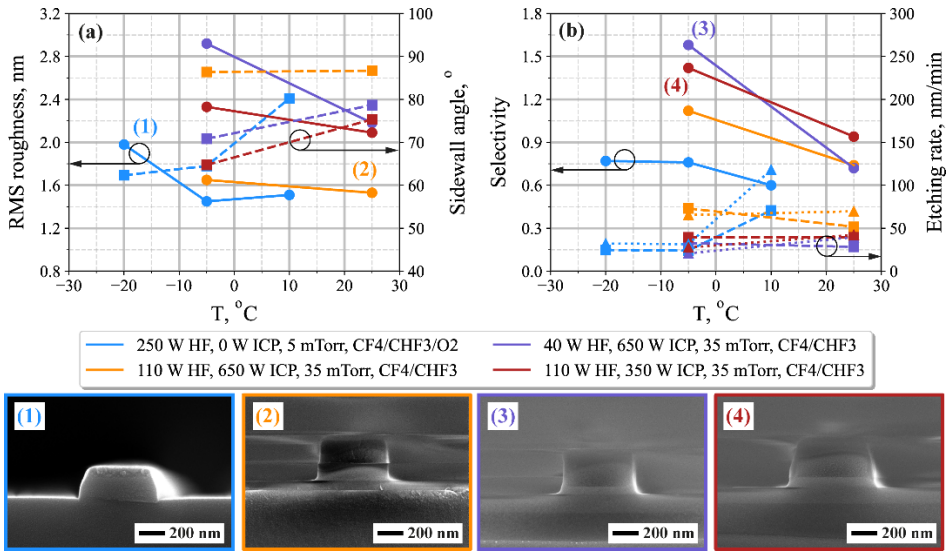

Fig. 8. Waveguide parameters versus platen temperature: (a) RMS sidewall roughness (solid lines) and sidewalls angle of inclination (dashed lines). (b) Etching selectivity, $Si_3N_4$ and resist etching rates (dashed and dotted lines, respectively). Cross-section of waveguides, fabricated with various etching processes are shown on SEM images.

It can be noticed that most etching processes, providing sidewall roughness decrease, have high etching rate, low selectivity and strong deviation of sidewalls angle from 90 °. For further $Si_3N_4$ ICP-RIE process optimization the variation of working pressure were carried out to increase etching selectivity. It was found, that RMS sidewall roughness dependence on working pressure has an extremum at 30 mTorr (Fig. 9a). This effect can be explained by the fact that with an increase in pressure, the amount of gas particles increases, while power fed in the discharge is sufficient to ensure bombardment. At pressures above 30 mTorr the input power becomes insufficient for effective bombardment, and reactive etching mechanism begins to dominate in the etching process. This is confirmed, firstly, by the appearance of a strong lateral etching during the process (SEM images inset in Fig. 9), and secondly, by a sharp increase in the etching selectivity to 2.2 (Fig. 9b). In order to achieve low RMS sidewall roughness while keeping high selectivity and vertical sidewalls we carried out additional optimization of our etching process (see Supplementary materials).

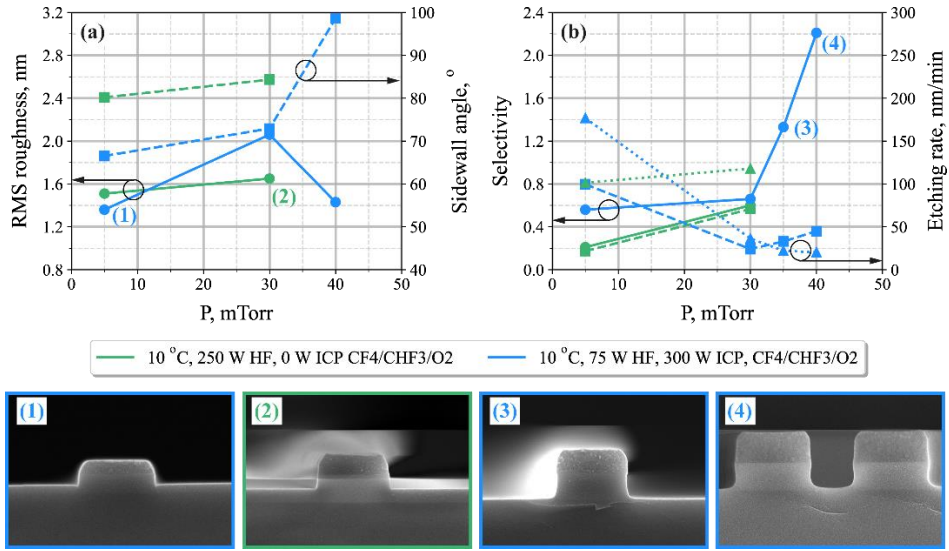

Fig. 9. Waveguides parameters versus working pressure: (a) RMS sidewall roughness (solid lines) and sidewalls angle of inclination (dashed lines). (b) Etching selectivity, $Si_3N_4$ and resist etching rates (dashed and dotted lines, respectively). Cross-section of waveguides, fabricated with various etching processes are shown on SEM images.

Finally, we optimized etching chemistry. We used $CF_4$ as the etchant. For slight passivation of sidewalls, we used $CHF_3$. Unlike common chemicals for etching $Si_3N_4$ like $CHF_3/O_2$ or $CHF_3/O_2/N_2$ [26, 43, 57], we removed oxygen as it limits selectivity to resist. Etching and passivation balance was achieved with optimized $CF_4$ and $CHF_3$ ratio at certain values of etching parameters. As a result of experiments, we developed ICP-RIE of $Si_3N_4$, which allows achieving almost vertical (89.5°) sidewalls with decrease in RMS sidewall roughness from 1.85 ± 0.21 nm to 1.08 ± 0.06 nm and to as low as 0.85 ± 0.06 nm with optimized e-beam lithography. The selectivity of developed ICP-RIE process was increased from 0.8 to 1.4.

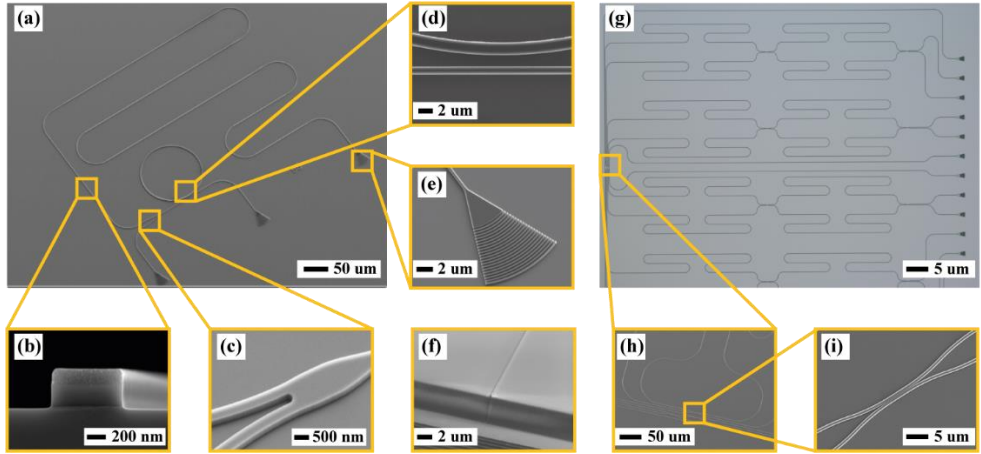

Fig. 10 Images of fabricated photonic circuits and structures: (a) Test photonic integrated circuit. (b) Single-mode submicron waveguide cross-section. (c) Top view of y-splitter. (d) Top view of microring resonator coupling section. (e) Top view of grating coupler. (f) Top view of taper coupler. (g) Multiport photonic integrated circuit. (h) Array of 6 tunable beam splitters (Mach-Zehnder interferometers with thermo-optic phase shifters). (i) Top view of directional coupler.

With the optimized technology, we are able to fabricate photonic integrated circuits consisting of single-mode submicron waveguides, y-splitters, microring resonators, grating and taper couplers, tunable beam splitters (Mach-Zehnder interferometers with thermo-optic phase shifters) and directional couplers. Images of fabricated photonic circuits and structures are shown in Figure 10.

## 4. Waveguide propagation losses modeling

To obtain propagation losses due to scattering one commonly uses the analytical approach by Payne and Lacey and the following expression [62]:

$$\alpha \left[\frac{dB}{cm}\right] = 4{,}34 \frac{\sigma^2}{k_0 \sqrt{2} d^4 n_1} gf, \qquad (1)$$

where $\alpha$ is the waveguide scattering loss in dB per unit length, $\sigma$ is the RMS deviation, $k_0$ is the free space wave vector, $d$ and $n_1$ are the waveguide half width and refractive index of $Si_3N_4$ core, respectively. Function $g$ is determined purely by the waveguide geometry, and $f$ is a function of correlation length and other parameters as defined by Payne and Lacey [63].

Payne-Lacey model provides rapid calculation of losses having good agreement with fully three-dimensional FDTD simulation and experimental results [64, 65]. However, it is impossible to estimate the influence of the roughness exponent $H$ on scattering losses with classical analytical approach. In this work, we propose the numerical solution for the Payne-Lacey model, which provides estimation of the influence of three main roughness parameters, measured with AFM or SEM – root mean square roughness $\sigma$, correlation length $\xi$ and roughness exponent $H$ (more information about our numerical approach could be found in Supplementary materials).

Figure 11 shows the calculation results for propagation losses from scattering for single-mode waveguides with 550 nm width at 925 nm wavelength.

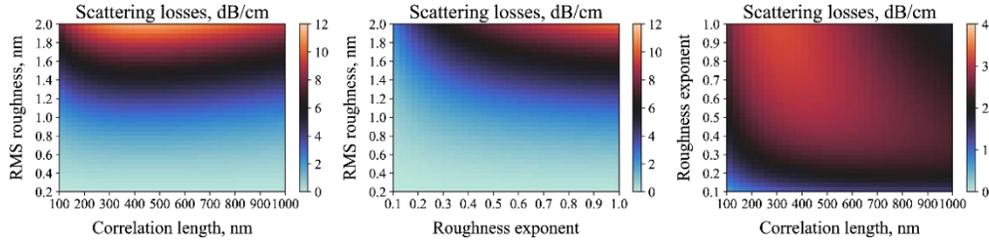

Fig. 11. Influence of scattering propagation losses from: (a) RMS roughness and correlation length at $H = 0.5$. (b) RMS roughness and roughness exponent at $\xi = 100$ nm. (c) roughness exponent and correlation length at $\sigma = 1.0$ nm.

Based on the numerical simulation results, the following conclusions were made.
1. High RMS roughness leads to increase of waveguide mode field interaction with sidewalls roughness, thus to higher scattering losses.
2. As correlation length tends to zero, mode field is less sensitive to changes in roughness at this frequency and is less scattered, even at large values of $\sigma$, which leads to low propagation losses. However, at very short wavelengths, low correlation length can bring higher impact. With an increase in $\xi$, an extremum is observed, indicating an increase in the interaction of the mode field with sidewalls and an increase in losses. As $\xi$ tends to infinity, implying no roughness, the loss tends to zero, since the frequency of the change in the roughness amplitude is also tends to zero.
3. At higher roughness exponent values, the mode field sensitivity to amplitude roughness is at maximum, leading to high scattering losses. However, when roughness exponent is low, that corresponds to high-frequency sidewall roughness, at our wavelength of interest the mode field is insensitive to such high-frequency profile variations, leading to low propagation losses, even at high RMS roughness. Same as correlation length, roughness exponent can have a higher impact on propagation losses at shorter wavelengths.

## 5. Optical characterization

The "cut-back" propagation loss analysis [66, 67] in fabricated $Si_3N_4$ waveguides has been performed using the optical characterization setup shown in Fig. 12a. The light from a continuous wave (CW) laser source at 925 nm wavelength was first coupled in single-mode fiber. TE-polarized light was formed with fiber polarization controller (FPC). Using 6-axis micrometer coupling stages, light from fiber was then coupled to waveguides through on-chip grating couplers. After propagating in photonic integrated circuit, the light was coupled out to power detector (PD). Propagation losses were measured on test photonic integrated circuits with various lengths of waveguides, fabricated with initial fabrication process, optimized e-beam lithography and optimized fabrication process (optimized e-beam lithography and ICP-RIE).

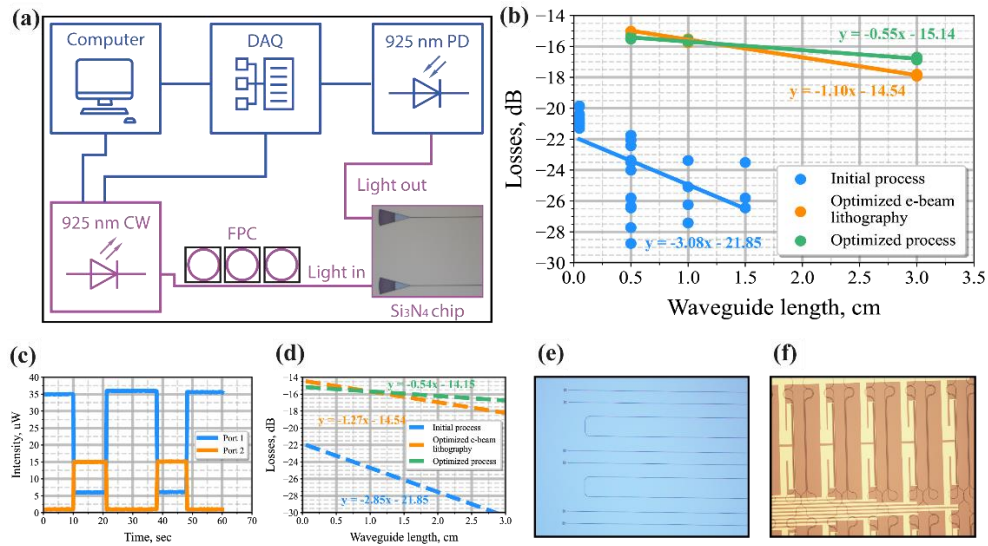

Fig. 12. Optical characterization of fabricated waveguides. (a) Optical characterization setup for "cut-back" propagation loss analysis in fabricated waveguides. (b) Measured propagation losses in fabricated $Si_3N_4$ waveguides with different length and fabrication process. (c) Demonstration of thermo-optic light switching with fabricated tunable beam splitters. (d) Theoretical scattering losses in fabricated waveguides, calculated with proposed numerical solution for the Payne-Lacey model. (e) Top view optical microscope image of fabricated structures (grating couplers and waveguides) for propagation losses measurement. (f) Top view optical microscope image of fabricated tunable beam splitters. CW: continuous wave laser, FPC: fiber polarization controller, PD: photodetector, DAQ: data acquisition system.

Figure 12b shows measured propagation losses in waveguides versus waveguide length. With the initial fabrication technology, a large spread of losses is observed at measured waveguides due to the presence of large stitches between the working fields and high sidewall roughness (blue line in Fig. 12b). As a result of e-beam lithography optimization, stitching error was greatly reduced, leading to decrease in coupling losses from 10.92 dB to 7.27 dB. Resist roughness improvement with optimized lithography shows reduction in propagation losses from 3.08 dB/cm to 1.10 dB/cm (orange line in Fig. 12b). Optimization of the ICP-RIE further reduced the RMS sidewall roughness of the waveguides, leading to propagation losses as low as 0.55 dB/cm (green line in Fig.12b). This result confirms that the main contribution to the propagation losses at 925 nm wavelength is made by scattering from waveguides bottom, sidewalls and upper surfaces roughness. Fabrication process optimization leads to roughness of waveguide sidewalls reduction by factor of 2, resulting in 6 times improvement in losses. Thermo-optic light switching with fabricated tunable beam splitters is shown on Figure 12c, demonstrating high quality and energy-efficient heaters fabrication process.

The results for scattering losses calculation with proposed numerical method are shown in Fig. 12d. Taking into account the roughness parameters measured on waveguides with initial fabrication process, the optimized e-beam lithography and the optimized fabrication process (for more information see Supplementary materials) we compared experimentally obtained propagation losses with calculated ones. The proposed numerical solution allows fairly accurate simulation of propagation losses in waveguides based on data from roughness parameters measurement, obtained with AFM or SEM with less than 10% deviation from experiment.

## 6. Conclusion

In this work we developed the fabrication process of low-loss silicon nitride photonic integrated circuits with in-depth study of e-beam lithography and inductively coupled plasma reactive ion etching steps for sidewalls roughness and profile angle improvement. The optimized multipass e-beam lithography process with reduced working field and warm development of conductive polymer results in stitching error reduction down to 1.04 ± 1.35 nm, vertical (89.5º) resist sidewalls and resist sidewall roughness as low as 1.17 ± 0.13 nm. The optimized ICP-RIE process based on the combination of etching and passivation effects of $CF_4$ and $CHF_3$ allows to reduce the final waveguide sidewalls roughness down to 0.85 ± 0.06 nm. With the optimized fabrication process, propagation losses in silicon nitride waveguides were reduced from 3.1 dB/cm to 0.55 dB/cm at 925 nm wavelength, which is the lowest demonstrated to date losses for single-mode submicron waveguides [15, 16, 38, 39, 40]. We proposed the enhanced Payne-Lacey analytical model for waveguide propagation losses modeling, which allows taking into account all the measured roughness parameters ($\sigma$, $\xi$ and $H$) for much more accurate loss estimation. Our result demonstrates the high potential for further improvements of propagation losses below 0.1 dB/cm by employing high optical quality $SiO_2$ cladding, chemical-mechanical polishing and multistep annealing of silicon nitride stack.

## Acknowledgments

Samples were made at the BMSTU Nanofabrication Facility (FMN Laboratory, FMNS REC, ID 74300) and measured at MSU Quantum Technology Centre.

## Disclosures

The authors declare no conflicts of interest.

## Data availability

The data that supports the findings of this study are available from the corresponding author upon reasonable request.

## Supplemental document

See Supplement 1 for supporting content.